\documentclass[sigconf]{acmart}

\usepackage{wrapfig}
\usepackage{graphicx}
\usepackage{url}

\copyrightyear{2018}
\acmYear{2018}
\setcopyright{acmcopyright}
\acmConference[CCS '18]{2018 ACM SIGSAC Conference on Computer \& Communications Security}{October 15--19, 2018}{Toronto, ON, Canada}
\acmBooktitle{CCS '18: 2018 ACM SIGSAC Conference on Computer \& Communications Security Oct. 15--19, 2018, Toronto, ON, Canada}
\acmPrice{15.00}
\acmDOI{10.1145/3243734.3264421}
\acmISBN{978-1-4503-5693-0/18/10}
\usepackage{booktabs} 
\setcopyright{rightsretained}

\acmArticle{4}
\acmPrice{15.00}

\editor{Jennifer B. Sartor}
\editor{Theo D'Hondt}
\editor{Wolfgang De Meuter}

\begin{document}
\title{Game Theory Meets Network Security}
\subtitle{A Tutorial}

\author{Quanyan Zhu}
\affiliation{%
  \institution{New York University}
  \streetaddress{5 MetroTech Center}
  \city{Brooklyn}
  \state{New York}
  \postcode{11201}
}
\email{qz494@nyu.edu}

\author{Stefan Rass}
\affiliation{%
  \institution{Universitaet Klagenfurt}
  \streetaddress{Universitaetsstr. 65-67, A-9020 }
  \city{Klagenfurt}
  \state{Austria}
}
\email{stefan.rass@aau.at}

\renewcommand{\shortauthors}{Q. Zhu and S. Rass}

\begin{abstract}
The increasingly pervasive connectivity of today's information systems brings up new challenges to security. Traditional security has accomplished a long way toward protecting well-defined goals such as confidentiality, integrity, availability, and authenticity. However, with the growing sophistication of the attacks and the complexity of the system, the protection using traditional methods could be cost-prohibitive. A new perspective and a new theoretical foundation are needed to understand security from a strategic and decision-making perspective. Game theory provides a natural framework to capture the adversarial and defensive interactions between an attacker and a defender. It provides a quantitative assessment of security, prediction of security outcomes, and a mechanism design tool that can enable security-by-design and reverse the attacker's advantage.  This tutorial provides an overview of diverse methodologies from game theory that includes games of incomplete information, dynamic games, mechanism design theory to offer a modern theoretic underpinning of a science of cybersecurity.  The tutorial will also discuss open problems and research challenges that the CCS community can address and contribute with an objective to build a multidisciplinary bridge between cybersecurity, economics, game and decision theory.

\end{abstract}

%
%



\maketitle


\section{Tutorial Description}


Contemporary information and communication technology evolves fast not only in terms of the level of sophistication but also regarding its diversity. The increasing complexity, pervasiveness, and connectivity of today's information systems brings up new challenges to security, and the cyberspace has become a playground for people with all levels of skills and all kinds of intention (positive and negative). With 24/7 connectivity having become an integral part of people's daily life, protecting information, identities, and assets has gained more importance than ever. 
Traditional security has accomplished a long way toward protecting well-defined goals such as confidentiality, integrity, availability, and authenticity (CIA+). Cryptography is a solid theoretic foundation for security which relies on the secrecy of cryptographic keys. However, for attackers who can steal full cryptographic keys as in advanced persistent threats (APTs) or social engineering attacks, the assumption of key secrecy fails, and they can penetrate the system. A new perspective and a new theoretical foundation are needed to capture scenarios where an attacker can completely compromise a system, and a defender aims to protect the system without the assumption of key secrecy.

%
%



Game-theoretic models are natural frameworks to capture the adversarial and defensive interactions between players \cite{zhu2018multi,rass2017physical,zhuang2010modeling,miao2018hybrid,farhang2014dynamic,manshaei2013game,zhu2013deployment,zhang2017strategic,horak2017manipulating,huang2017large}.    Game theory can provide a quantitative measure of the quality of protection with the concept of Nash equilibrium where both defender and an attacker seek optimal strategies, and no one has an incentive to deviate unilaterally from their equilibrium strategies despite their conflict for security objectives. The equilibrium concept also provides a quantitative prediction of the security outcomes of the scenario the game model captures.  With the quantitative measures of security, game theory makes security manageable beyond the strong qualitative assurances of cryptographic protections. Extending this approach to mechanism design provides system designers freedom to shift the equilibrium and the predicted outcomes toward ones that are favored by the defender or the system designer via an elaborate design of the game structure.

%

For more than a decade now, the interest in the field has proliferated, and game- and decision theory has become a systematic and well proven powerful theoretic underpinning of today's security research. Somewhat different from standard security definitions, game- and decision theory adopts a different and more economic viewpoint: security is not the absence of threats, but the point where attacking a system has become more expensive than not attacking. Starting from a game- and decision-theoretic root thus achieves the most elegant type of self-enforcing security, by analyzing and creating incentives to encourage actively honest behaviors rather than preventing maliciousness. At the same time, the economic approach to security is also essential as it parallels the evolution of today's attackers. Cybercrime has grown into a full-featured economy, maintaining black markets, supply chains, and widely resembling an illegal counterpart of the of the official software market. Traditional security remains an important foundation to tackle the issue from below, but game- and decision theory offers a top-down view by adopting the economic and strategic view of the attackers too, and as such complements purely technological security means. The optimum is achieved when both routes are taken towards meeting in the middle, which is what game and decision theory aims to achieve.

The objective of this tutorial is to introduce diverse methodologies from
game theory that include mechanism design, incentive analysis,
decision-making under incomplete information, and dynamic games to provide
solid underpinnings of a science of cybersecurity. The tutorial will be
organized to connect different classes of games with different sets of
security problems. For example (1) Stackelberg and multi-layer games for
proactive defense
\cite{pawlick_stackelberg_2016,zhu2013game,zhu2013deployment,zhu2013hybrid,zhu2012interference,clark2012deceptive,zhu2012game,zhu2012deceptive,zhu2010stochastic},
(2) network games for cyber-physical security that deals with critical
infrastructure protection and information assurance
\cite{xu2017secure,xu_game-theoretic_2017,xu_cross-layer_2016,xu2015cyber,huang2017large,chen2017dynamic,miao2018hybrid,yuan2013resilient,Rass&Zhu2016,Rass.2017b},
(3) dynamic games for adaptive defense for network security
\cite{zhu2010dynamic,zhang2017strategic,huang2018gamesec,huang2018PER,pawlick2015flip,farhang2014dynamic,zhu2009dynamic,zhu2010network,zhu2010heterogeneous},
(4) mechanism design theory for economics of network security that
investigates resource allocation methodologies
\cite{chen_security_2017,zhang_bi-level_2017,zhang_attack-aware_2016,casey2015compliance,hayel2015attack,hayel2017epidemic,zhu2012guidex,zhu2012tragedy,zhu2009game},
and (5) game-theoretic analysis of cryptographic concepts, such as perfectly
confidentiality and authentication (in classical and quantum networks)
\cite{Rass&Schartner2009,Rass&Schartner2010,Rass&Schartner2011c,Rass&Schartner2011},
network design and -provisioning
\cite{Rass&Wiegele&Schartner2010,Rass&Rainer2015,Rass&Rainer2013,Rass2011,Rass2013c,Rass2014}
and quantitative security risk management
\cite{Rass2015b,Rass2015c,Rass.20171102,Rass.2018,Rass.2017d,Rass.2018b,Rass&Konig&Schauer2015b,Rass&Konig&Schauer2016,Rass&Konig&Schauer2017,Rass&Konig2016,Rass&Rainer2014,Rass&Schartner&Wigoutschnigg2010}

From the perspective of cybersecurity,  the topics of this tutorial will cover recent applications of game theory to several emerging topics such as
cross-layer cyber-physical security \cite{miao2018hybrid,zhu2018multi,pawlick2017strategic,chen2017security,zhu2015game,xu2017secure},
cyber deception \cite{pawlick2018modeling,zhang2017strategic,horak2017manipulating,pawlick2017game,pawlick2015deception,zhuang2010modeling},
moving target defense \cite{zhu2013game,jajodia2011moving,maleki2016markov},
critical infrastructure protection \cite{chen2017dynamic,rass2017physical,huang2017large,pawlick2017proactive,chen_interdependent_2016,hayel2015resilient,huang2018PER},
adversarial machine learning \cite{zhang2018game,wang2017detection,zhang2017game,pawlick_stackelberg_2016,pawlick_mean-field_2017},
insider threats \cite{casey_compliance_2016,casey2015compliance},
and cyber risk management \cite{zhang2017bi,hayel2017epidemic,fung2016facid,chen2018security}. The tutorial will also discuss open problems and research challenges that the CCS community can address and contribute. With the objective to build a multidisciplinary bridge between cybersecurity, economics, game and decision theory, this tutorial will review basic concepts and provide an overview of recent advances in the field to CCS community with the hope to establish a community interest in the science of security and cross-disciplinary researches.

%
The potential audience includes researchers from academia and industry, including PhD and graduate students. Some background in network security and knowledge of basic optimization and data science is helpful but not necessary. The tutorial takes 1.5 hours.

\section{Author Biography}

\begin{wrapfigure}{r}{0.14\textwidth}
  \begin{center}
  \vspace{-12mm}
   \includegraphics[width=0.13\textwidth]{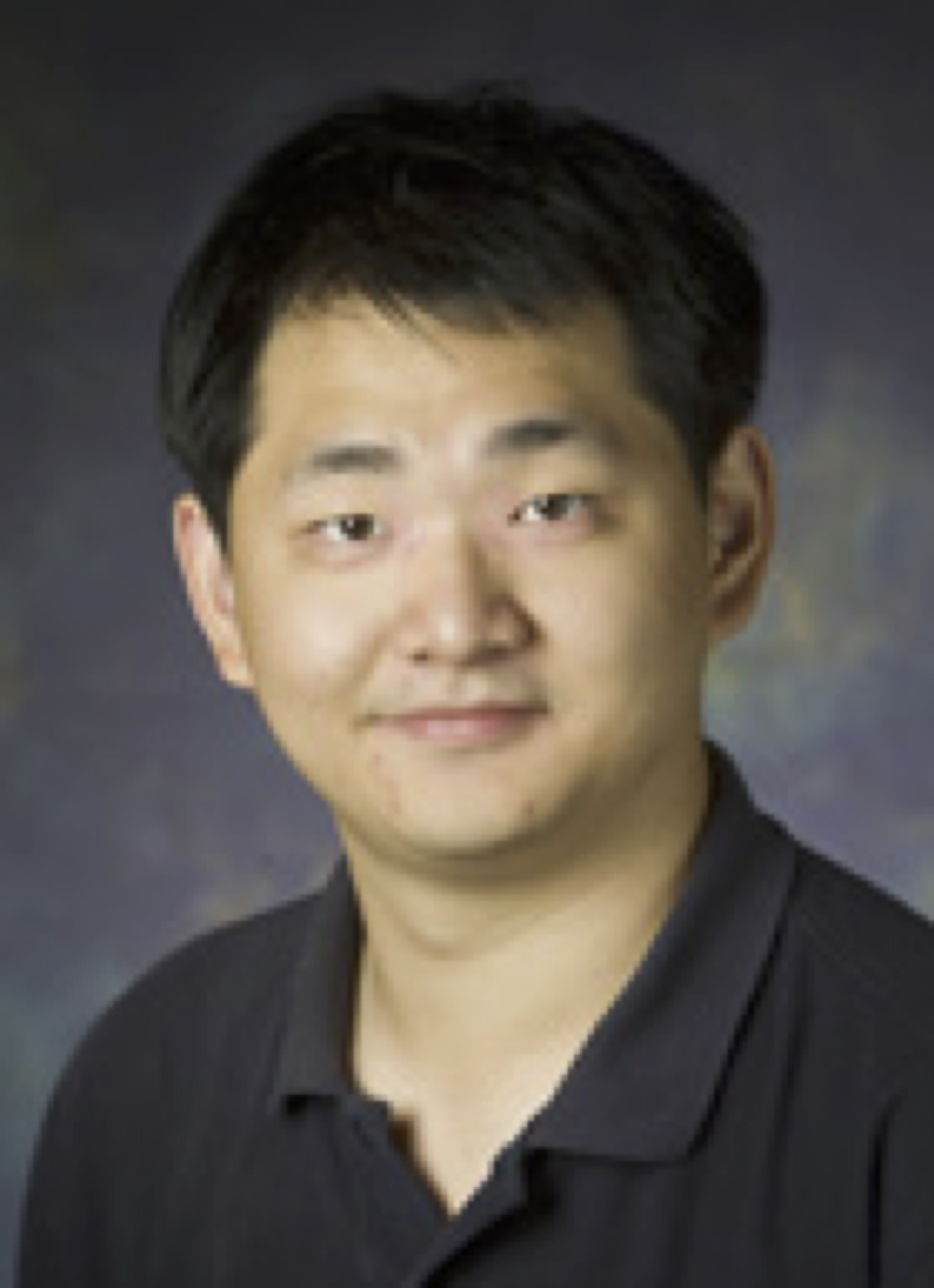}
  \end{center}  
\end{wrapfigure}

Quanyan Zhu received B. Eng. in Honors Electrical Engineering from McGill University in 2006, M.A.Sc. from University of Toronto in 2008, and Ph.D. from the University of Illinois at Urbana-Champaign (UIUC) in 2013. After stints at Princeton University, he is currently an assistant professor at the Department of Electrical and Computer Engineering, New York University. He is a recipient of many awards including NSERC Canada Graduate Scholarship (CGS), Mavis Future Faculty Fellowships, and NSERC Postdoctoral Fellowship (PDF). He spearheaded and chaired INFOCOM Workshop on Communications and Control on Smart Energy Systems (CCSES), and Midwest Workshop on Control and Game Theory (WCGT). His current research interests include resilient and secure interdependent critical infrastructures, energy systems, cyber-physical systems, and cyber-enabled sustainability. He is a recipient of best paper awards at 5th International Conference on Resilient Control Systems, and 18th International Conference on Information Fusion. He has served as the general chair of the 7th Conference on Decision and Game Theory for Security (GameSec) in 2016 and International Conference on NETwork Games, COntrol and OPtimisation (NETGCOOP) in 2018.
Website: \url{http://wp.nyu.edu/quanyan}

\begin{wrapfigure}{r}{0.14\textwidth}
  \begin{center}
  \vspace{-4mm}
   \includegraphics[width=0.13\textwidth]{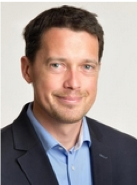}
  \end{center}  
\end{wrapfigure}

\smallskip

Stefan Rass graduated with a double master degree in mathematics and computer science from the Universitaet Klagenfurt in 2005. He received a PhD degree in mathematics in 2009, and habilitated on applied computer science and system security in 2014. His research interests cover decision theory and game-theory with applications in system security, as well as complexity theory, statistics and information-theoretic security. He authored numerous papers related to security and applied statistics and decision theory in security. He (co-authored) the book ``Cryptography for Security and Privacy in Cloud Computing", published by Artech House, and edited the Birkh\"auser Book ``Game Theory for Security and Risk Management: From Theory to Practice" in the series on Static \& Dynamic Game Theory: Foundations \& Applications. He participated in various nationally and internationally funded research projects, as well as being a contributing researcher in many EU projects and offering consultancy services to the industry. Currently, he is an associate professor at the AAU, teaching courses on algorithms and data structures, theoretical computer science, complexity theory, security and cryptography.	
Website: \url{https://www.syssec.at/en/team/rass}

\bibliographystyle{acm}

\bibliography{MyPapers-BIBTEX-2018,additional-refs}

\end{document}